\documentclass{osa-article}

\journal{osajournal}
\usepackage{threeparttable}
\articletype{Research Article}
\begin{document}

%\title{Production of Watt-level SHG from a single-frequency tunable VECSEL in the region 910-928 nm}
%\title{Watt-level blue light from a single-frequency tunable VECSEL for precision spectroscopy, laser cooling and trapping of strontium and cadmium atoms}
\title{Watt-level blue light for precision spectroscopy, laser cooling and trapping of strontium and cadmium atoms}

%\author{Author One\authormark{1}, Author Two\authormark{2,3}, and Author Three\authormark{1,4*}}
\author{Jonathan N. Tinsley,\authormark{1} Satvika Bandarupally,\authormark{1} Jussi-Pekka Penttinen,\authormark{2,3} Shamaila Manzoor,\authormark{1}  Sanna Ranta,\authormark{2,3} Leonardo Salvi,\authormark{1} Mircea Guina,\authormark{2,3} and Nicola Poli\authormark{1,4*}}

\address{\authormark{1}Dipartimento di Fisica e Astronomia and LENS - INFN Sezione di Firenze, Universit\`{a} degli Studi di Firenze, Via Sansone 1, 50019 Sesto Fiorentino, Italy\\
\authormark{2}Optoelectronics Research Centre, Tampere University, 33720 Tampere, Finland\\
\authormark{3}Vexlum Ltd, Korkeakoulunkatu 3, 33720 Tampere, Finland\\
\authormark{4}Istituto Nazionale di Ottica del Consiglio Nazionale delle Ricerche (INO-CNR), 50019 Sesto Fiorentino, Italy}
%\authormark{2}Publications Department, The Optical Society (OSA), 2010 Massachusetts Avenue NW, Washington, DC 20036, USA}

\email{\authormark{*}nicola.poli@unifi.it} %% email address is required



\begin{abstract}
High-power and narrow-linewidth laser light is a vital tool for atomic physics, being used for example in laser cooling and trapping and precision spectroscopy. Here we produce Watt-level laser radiation at 457.49~nm and 460.86~nm of respective relevance for the cooling transitions of cadmium and strontium atoms. This is achieved via the frequency doubling of a kHz-linewidth vertical-external-cavity surface-emitting laser (VECSEL), which is based on a novel gain chip design enabling lasing at >~2~W in the 915-928~nm region. Following an additional doubling stage, spectroscopy of the $^1S_0\to{}^1P_1$ cadmium transition at 228.89~nm is performed on an atomic beam, with all the transitions from all eight natural isotopes observed in a single continuous sweep of more than 4~GHz in the deep ultraviolet. The absolute value of the transition frequency of $^{114}$Cd and the isotope shifts relative to this transition are determined, with values for some of these shifts provided for the first time.
\end{abstract}

%%%%%%%%%%%%%%%%%%%%%%%%%%  body  %%%%%%%%%%%%%%%%%%%%%%%%%%

\section{Introduction}
Alkaline-earth and alkaline-earth-like atoms, such as strontium and cadmium, have attracted considerable recent interest and research in the field of high-precision metrology and inertial sensing due to a number of favorable characteristics. For example, the presence of two valence electrons provides access to narrow intercombination transitions which have been utilized for e.g. high-precision optical atomic clocks and for atom interferometry. Strontium has received considerable attention, especially in the atomic clock community~\cite{Ushijima_2015,Campbell_2017,Oelker_2019}, but also more recently for atomic interferometry, with recent demonstrations on the narrow $^1S_0\to{}^3P_1$ intercombination~\cite{delAguila_2018,Rudolph_2020} and ultra-narrow $^1S_0\to{}^3P_0$ clock transitions~\cite{Hu_2017,Hu_2019}, as well as being suggested as the test species for both ground-based~\cite{MAGIS_2018,AION_2020} and satellite missions~\cite{SAGE_2019,AEDGE_2020} for gravitational wave and dark matter detection. Cadmium is presently less well developed, though it has attracted renewed interest due to its low susceptibility to background blackbody radiation~\cite{Yamaguchi_2019,Porsev_2020}. Strontium and cadmium also possess favorable features which would be beneficial when operating as a dual-species atom interferometer used to search for possible violations of the weak equivalence principle or other extensions to general relativity~\cite{Tinsley_2019}. The frequencies of all of the $^1S_0\to{}^1P_1$, $^1S_0\to{}^3P_0$ and $^1S_0\to{}^3P_1$ transitions of strontium and cadmium are close to being in a 1:2 ratio (Sr:Cd), allowing for the momentum-transfer and Rabi frequencies in a dual-Bragg interferometer to be almost equal, which is important for ensuring common-mode rejection of noise and other systematic effects~\cite{Barrett_2015}. Additionally, the presence of both fermionic and bosonic isotopes of both atoms allows for potential spin-gravity couplings to be investigated~\cite{Tarallo_2014} and using a Sr-Cd dual interferometer would provide good sensitivity to both standard model extensions and dilaton models~\cite{Damour_2012,Hohensee_2013}.

%High-power, continuous-wave, single-mode and frequency lasers are one of the vital tools of contemporary optical, atomic and molecular research. Often, however, the required properties of narrow linewidth, high power and single-mode operation are not easily fulfilled by a single system and certain wavelength regions remain relatively poorly covered due to technical challenges. One example of such a relative paucity is the near infrared (NIR) region around 900-940~nm~\cite{Rota-Rodrigo_2017}. The impact of this lack of coverage is compounded by an unfortunate lack of similar sources in the corresponding frequency-doubled region (450-470~nm). Simple and affordable sources capable of generating high-power light in the NIR and, via second harmonic generation (SHG), in the blue portion of the visible spectrum, are therefore of great interest for a variety of fields of research and industrial applications, \textcolor{red}{including...}.

Crucial to enabling all of these experiments is the laser cooling, trapping and manipulation of atomic samples. A key transition in this respect is the $^1S_0\to{}^1P_1$ transition, with respective wavelengths of 460.86~nm and 228.89~nm for strontium and cadmium. For example, the $^1S_0\to{}^1P_1$ transition is key in the preparation of cold strontium samples, being employed in Zeeman slowers and magneto-optical traps~\cite{Courtillot_2003,Xu_2003}, as well as being used to drive Bragg transitions in atomic interferometers~\cite{Mazzoni_2015}. Similarly for cadmium the deep ultraviolet (DUV) transition at 228.89~nm is used for cooling and trapping, with the DUV produced by frequency-doubling blue light~\cite{Yamaguchi_2019,Brickman_2007,Kaneda_2016}. There are, however, a limited number of options for producing laser light in the region 450-470~nm with the fulfilment of all of the required properties of narrow linewidth, high power and single-mode operation with good mode quality. The impact of this lack of coverage is compounded by an unfortunate lack of similar sources in the corresponding near-infrared (NIR) region (900-940~nm), which could otherwise be used as a master source for the production of the necessary blue light via second-harmonic generation (SHG).

%One area which would benefit from improved sources in this region, especially following SHG, is the laser cooling, trapping and manipulation of atomic samples. For example, the $^1S_0\to{}^1P_1$ transition of strontium at 460.86~nm is a key transition in the preparation of cold strontium samples, being employed in Zeeman slowers and magneto-optical traps~\cite{Courtillot_2003,Xu_2003}, as well as being used to drive Bragg transitions in atomic interferometers~\cite{Mazzoni_2015}. All of these applications are aided by narrow-linewidth, high-power lasers at this wavelength. Furthermore, such laser sources can also serve as a fundamental source for the generation of deep ultraviolet light (DUV), for example the 228.89~nm light necessary for the $^1S_0\to{}^1P_1$ transition of cadmium, which is similarly employed for laser cooling and trapping~\cite{Brickman_2007,Kaneda_2016,Yamaguchi_2019}.

Solutions for producing blue light, either directly or via SHG, with all of these properties do exist. For example, there are high-power, tunable sources in the 900-940~nm region based upon e.g. Ti:sapphire lasers~\cite{Fengqin_2015} or neodymium-doped fibre~\cite{Rota-Rodrigo_2017}, the latter of which can produce Watt-level light in the region 915-937~nm. These systems, however, are complex and typically necessitate a large footprint and expense. Therefore, experiments often rely on tapered amplifier systems which, while capable of producing usable power in the region 500-1000~mW following SHG~\cite{Barbiero_2020}, suffer from bad ageing effects, moderate intensity noise and, especially, poor and variable beam quality, limiting their practical use. More recently, efforts at producing blue laser light directly from diodes have advanced and very recently a commercial system for producing 1~W directly at 460.86~nm has been made available. To achieve competitive powers these systems typically require a master-and-slave configuration with an extended-cavity diode laser (ECDL) used to inject a second diode, with long-term operation limited by the maintenance of good injection and mode quality suitable for only 65\% coupling to a single-mode fibre\cite{Schkolnik_2020}. Due to the relatively new nature of this technology there remain some open questions, however, regarding for example the typical lifetime of the diodes.

In this paper, we instead develop a vertical-external-cavity surface-emitting laser (VECSEL)~\cite{Kuznetsov_1997}, which is tunable in the region between around 910-928~nm, with Watt-level output. VECSELs have the advantage of producing high-power, high-brightness, narrow-linewidth and low-noise tunable single-mode light in a relatively simple configuration which minimizes the mechanical footprint of the laser (19 $\times$ 32~cm$^2$). With respect to its low-noise properties, VECSELs are so-called class-A lasers, with dynamics governed by the photon lifetime and therefore don't exhibit relaxation oscillations. Moreover, they also don't suffer from the amplified spontaneous emission present in amplified semiconductor laser diode systems. Due to this combination of desirable features, VECSELs have been employed recently in quantum technology applications, for example in the demonstration of full-scale functionality for the manipulation of Mg~\cite{Burd_2016} and Be~\cite{Burd_2020} ions, and have been recognised as an attractive solution for other ion and neutral atom systems, for example pertaining to the intercombination transition of strontium~\cite{Moriya_2020} and for the cooling and trapping of cadmium~\cite{Kaneda_2016}. Here we use a common design for an external resonant frequency-doubling, bow-tie cavity capable of producing up to 1~W of stable SHG, which when combined with a suitable doubling crystal effectively extends the tunability and performance of the VECSEL itself into the blue region of the visible spectrum. Specifically, we have developed a system which is suitable to act as a master source for the laser cooling and trapping on the $^1S_0\to{}^1P_1$ transition of both strontium and cadmium. We anticipate good long-term operation for this system, as the VECSEL in the 900-nm range is made with a similar reliable material system to that used in 980-nm diode pump lasers for telecommunication fibre amplifiers, and the optical pumping of the VECSEL is performed with a reliable 808-nm diode, commonly employed to pump neodymium-doped yttrium aluminium garnet (Nd:YAG) lasers.

In addition, we use the VECSEL as a master laser to perform spectroscopy upon a beam of atomic cadmium, identifying the $^1S_0\to{}^1P_1$ transitions from all eight naturally occurring isotopes and measuring their relative and absolute transition frequencies. Whilst absolute measurements of the $^1S_0\to{}^1P_1$ transition of other atoms has received attention~\cite{Laupretre_2020}, measurements on cadmium are more limited and often decades old~\cite{Burns_1956,Moore_1958,NIST_Spectra}. Moreover, some of these data were determined from interferometers operating in air and therefore requiring the refractive index at the transition wavelength of 228.89~nm to be well known. Whilst empirical formulae exist for calculating the refractive index of air as a function of wavelength~\cite{Peck_1972}, the refractive index also depends upon other environmental factors, for example the ambient temperature, pressure and humidity.

Complete measurement of the isotope and hyperfine shifts of the $^1S_0\to{}^3P_1$ intercombination transition of cadmium at 326~nm have been performed~\cite{Maslowski_2009}, but comparable data for the $^1S_0\to{}^1P_1$ transition is lacking, being only available in fragmentary form~\cite{Kelly_1961} or inferable from the published data of the frequency dependence of trapping in magneto-optical traps~\cite{Brickman_2007,Kaneda_2016}. Furthermore, precision measurement of isotope shifts provides important data for theoretical calculations of atomic structure~\cite{Miyake_2019}, as well as presenting the opportunity to search for and constrain potential extensions to the standard model of elementary particles and interactions, such as new long-range or Higgs-like forces~\cite{Delaunay_2017,Berengut_2018}.

\section{Laser Source Setup}
\subsection{Master Laser: NIR VECSEL}
\begin{figure}[t]
\centering\includegraphics[width=0.91\textwidth]{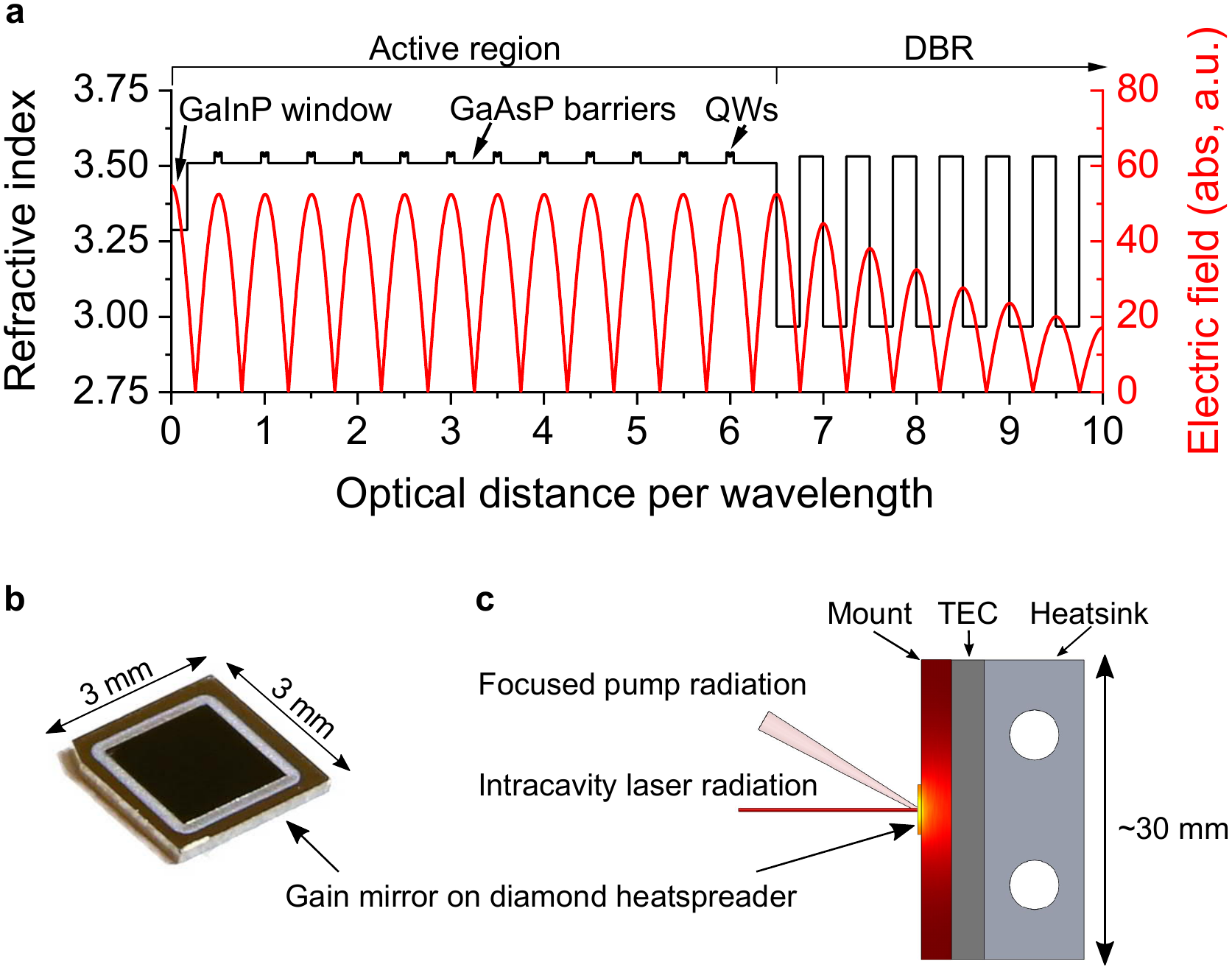}
\caption{a) VECSEL gain mirror structure as depicted with refractive index (black line) on left-hand axis. The simulated standing-wave electric field modulus (red line) is a critical design parameter for the thickness of the layers and is plotted on the right-hand axis for reference. b) Photograph of the processed flip-chip gain mirror on a $3 \times 3 \times 0.3$~mm$^3$ diamond heatspreader. c) Schematic of the gain mirror cooling assembly incorporating the gain mirror on the heatspeader, a cooling mount, a thermoelectric element (TEC), and a water-cooled heatsink.\label{fig:figureVECSELChip}}
\end{figure}
The fundamental laser source is a VECSEL based on a novel gain mirror designed for high-power emission in the 900-930-nm emission band, operating with an 808-nm pump diode source. The monolithic semiconductor multi-layer structure is depicted in Fig.~\ref{fig:figureVECSELChip}a (left axis) and was grown on a GaAs substrate using solid-source molecular beam epitaxy (MBE). The structure incorporates an active region with GaInP window (first-grown layer) and multiple GaInAs quantum wells (QWs) embedded in GaAsP barriers, and a distributed Bragg reflector (DBR) formed by alternating $\lambda$/4-thick layers of AlAs and GaAs. As the carrier confinement of a single low-indium GaInAs QW near the GaAs band-gap $\sim$870~nm is low, a large number of QWs are used to mitigate possible carrier leakage. A total of 24 (12~$\times$~2) QWs are placed in the antinodes of the simulated cavity standing wave; this is a relatively high number compared to typical gain structure operating at the 1-$\mu$m wavelength range (approx. double). The choice is governed by the need to ensure a good carrier collection in the QWs and hence sufficient single-pass gain. For the same reason GaAsP barriers are used~\cite{Zhang_2009}, instead of the more commonly used GaAs barriers, to increase the QW confinement while maintaining a good level of absorption for the 808-nm pump photons and ensuring good optical material quality.

The grown and diced gain mirror is flip-chip bonded~\cite{Kuznetsov_1997} from the mirror back surface onto a high thermal conductance $3 \times 3 \times 0.3$~mm$^3$ diamond heatspreader, after which the GaAs substrate is removed with selective wet-etching, thus revealing the GaInP window/etchstop layer (see photograph in Fig.~\ref{fig:figureVECSELChip}b). The new air/semiconductor surface is antireflection coated using ion-beam sputtering (IBS) to minimize the reflection of the focused pump light (AOI $\sim$30$^\circ$), to optimize the reflection of the intracavity light (AOI 0$^\circ$), and to protect the semiconductor surface. The remaining semiconductor structure on diamond is only $\sim$5~$\mu$m thick and serves as a disk-shaped gain element with an $\sim$80 aspect ratio compared with the used $\sim$400~$\mu$m pump diameter. The disk-shaped geometry enables efficient heat extraction of the pump laser load via the heatspeader diamond, to an actively TEC-stabilized cooling mount, and further to a heatsink cooled with a low-vibration chiller, which helps maintain the whole laser housing at a near constant temperature.

\begin{figure}[t]
\centering\includegraphics[width=\textwidth]{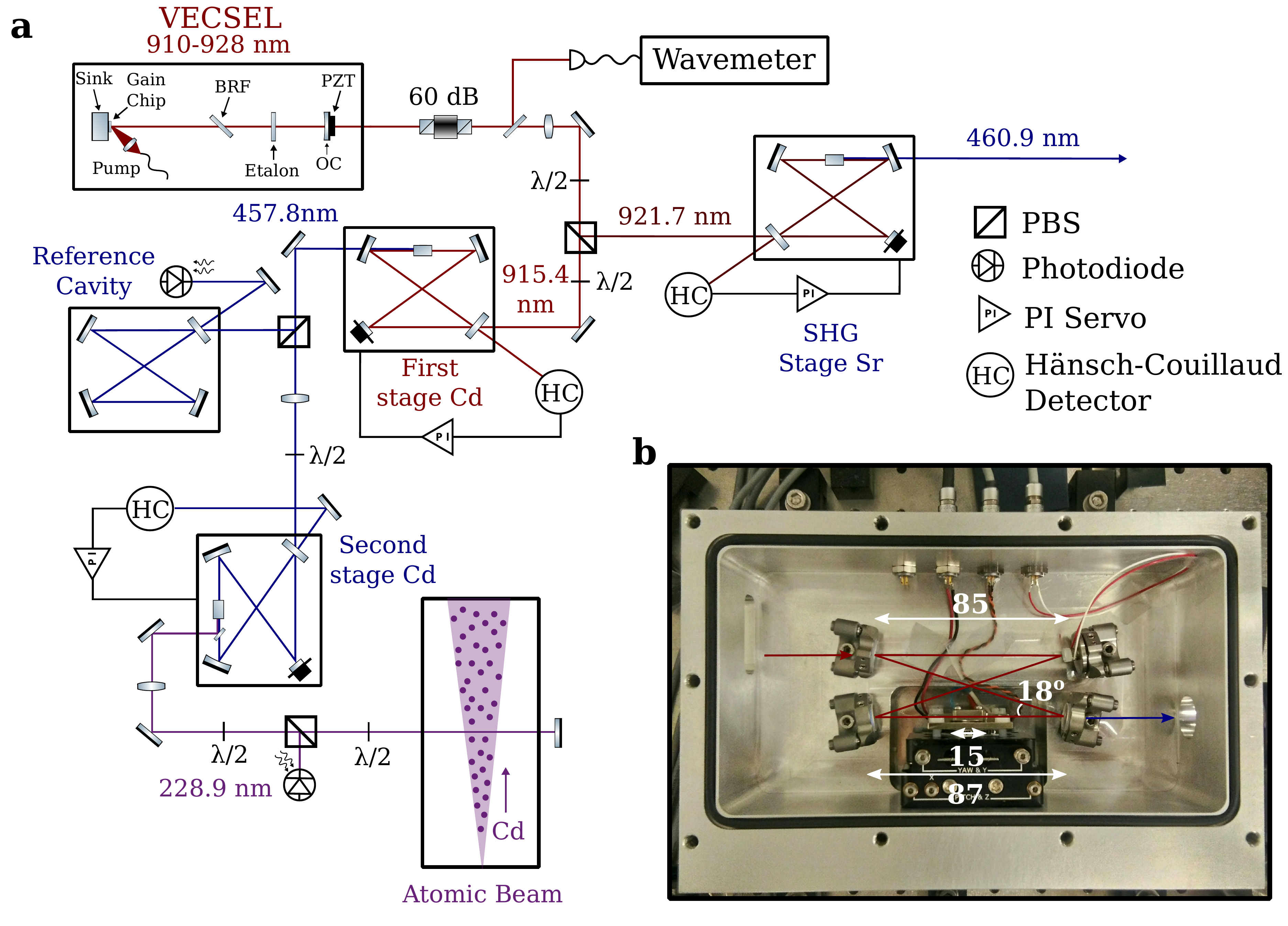}
\caption{a) A tunable VECSEL is employed as a master laser source, with the wavelength selectable via the combination of the angle of a birefrigent filter (BRF), the temperature of an etalon, and the position of the outcoupling mirror (OC). After passing through an optical isolator, the output of a VECSEL is mode-matched to a resonant bow-tie cavity with an LBO crystal to produce blue light for experiments with either strontium or cadmium. When performing experiments with Cd, a second Brewster-cut BBO frequency-doubling cavity is employed. The DUV output of this second doubling stage is used to perform spectroscopy on a beam of atoms whilst a reference cavity monitors the change in frequency. For more details, see the text. b) A photograph of the SHG cavity without the lid, showing the cavity dimensions in mm.\label{fig:figureSetup}}
\end{figure}

A $\sim$12-cm long air-cavity, similar to~\cite{Burd_2016}, is formed between the gain mirror and an external curved output coupling mirror, ensuring that the pump spot diameter and the simulated fundamental mode diameter are matched. A temperature-stabilized etalon and birefringent filter are inserted inside the cavity to enable single-frequency operation and tuning (see Fig.~\ref{fig:figureSetup}a). The cavity is placed inside a sealed and temperature-stabilized casing for stable operation in changing atmospheric and temperature conditions. The casing is further purged with dry nitrogen to prevent potential deleterious effects from absorption by remaining atmospheric water vapor. Selection of the optical frequency is coarsely performed via the temperature control of the etalon and the temperature and angular control of the birefringent filter. Continuous frequency control across the full $\sim$1.2 GHz free-spectral range and locking of the laser frequency is enabled by a piezoelectric transducer (PZT) on which the output coupling mirror is mounted, providing a measured bandwidth of $\sim$50~kHz.

\subsection{SHG Production}
The output of the VECSEL is directed first through a 60~dB optical isolator before being mode-matched with a single biconvex, spherical lens to a frequency-doubling, bow-tie cavity which uses a lithium triborate (LiB$_3$O$_5$) crystal as the doubling medium (Fig.~\ref{fig:figureSetup}), the main design for which is inspired by a system at 399~nm~\cite{Pizzocaro_2014}. Two separate cavities, which share a common geometrical and mechanical and optical design, are installed, allowing for the fast switching between producing light for either strontium-based or cadmium-based experiments, with the power splitting controlled by a polarising beamsplitter and half-wave plate combination. The cavity geometry is designed to work across the range of the output of the VECSEL~\cite{Freegarde_2001}, with the only substantial difference for switching between the two blue wavelengths being the necessary phase matching angle of the crystal. The reflectivities of the mirrors are suitably high for the wavelengths required for both strontium and cadmium and are thus interchangeable.

An outline of the SHG cavity parameters and a photograph of the opened cavity are shown in Figs.~\ref{fig:figureSetup}a and~b, with distances kept small so as to minimize the size and footprint. Considering the case at 460.86~nm for strontium, two mirrors of radius of curvature 75~mm and with an angle of incidence of 9$^{\circ}$, provide a focus of 40~$\mu$m and 42~$\mu$m, in the tangential and sagittal dimensions respectively, at the centre of the crystal, with the calculated Boyd-Kleinman factor of 0.28 near to the maximum value for this wavelength and crystal length~\cite{Boyd_1968,Freegarde_1997}. Two plane mirrors fold the beam and support a secondary waist of 220~$\mu$m and 244~$\mu$m at their mid-point, again in the tangential and sagittal dimensions respectively. Light is coupled into this cavity via one of these mirrors, which has a transmission of 1\%, selected heuristically so as to optimize the power of the SHG~\cite{Pizzocaro_2014}. The total optical path length is 362~mm and the free-spectral range 829~MHz. The optical finesse of this cavity is around 400, as measured from the linewidth and by inducing sidebands at 10~MHz with an electro-optical modulator to act as frequency reference. The difference to these values when considering the case of the cadmium-specific 457.49~nm are negligible.

The cavity is monolithic in design for enhanced thermal and mechanical stability, being produced from a single block of aluminium and with the mirror mounts being screwed directly into this bulk material. Good phase matching of the crystal is ensured by mounting it on a five-axis aligner (Thorlabs) and providing active temperature control via a thermoelectric cooler. One of the plane mirrors is mounted onto a PZT which controls the optical path length of the cavity and allows it to be stabilized via the H\"ansch-Couillaud locking method~\cite{Hansch_1980}. This mirror is only 6.3~mm in diameter, compared to a diameter of 12.7~mm for the other mirrors, so as to minimize the mechanical load and thereby extend the PZT control bandwidth, with a calculated resonance frequency of order 100~kHz.

Finally, the whole cavity can be placed under vacuum with the lid being sealed by a rubber O-ring and a valve. Without this vacuum element, it is significantly more difficult to achieve a strong and stable lock due to the high circulating powers leading to localized heating within the cavity. The associated change in optical path length of the cavity, arising from the negative thermo-optic coefficient, means that the symmetry with respect to a change of the fundamental laser frequency or optical path length of the cavity is broken. For example, when changing the frequency with the PZT of the cavity, the sweep is effectively slowed when the PZT is pulling and vice versa when the PZT is pushing~\cite{Kaneda_2016}. Operating the cavity under vacuum removes this problem in addition to providing some protection from acoustic noise.

\section{Optical Characterization}
%\subsection{Characterisation of the VECSEL}
\begin{figure}[t]
\centering\includegraphics[width=\textwidth]{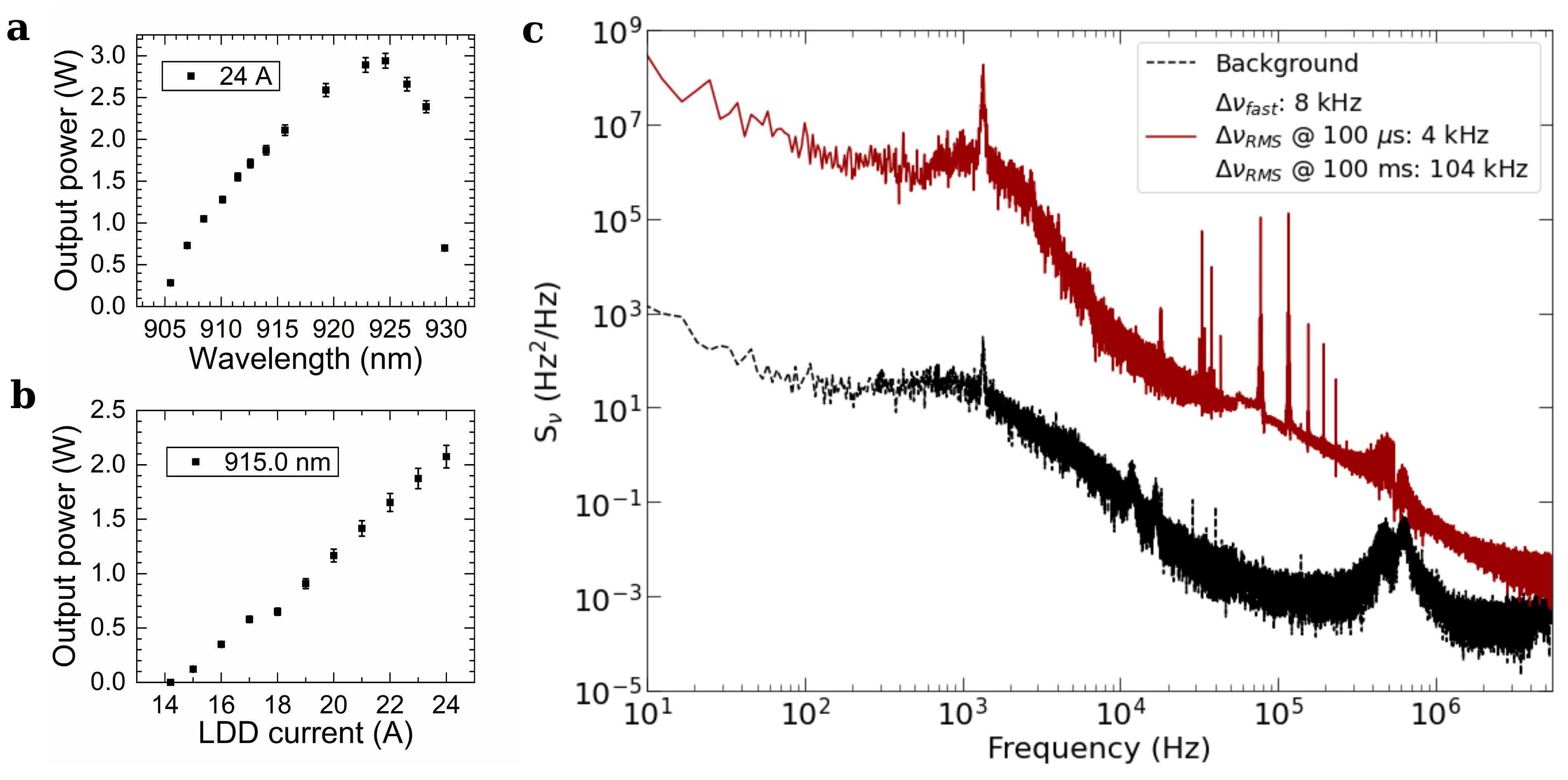}
\caption{a) Power as a function of wavelength with 24~A of pump laser current, close to saturation. b) The power output of the VECSEL at 915~nm as a function of the pump laser current. c) Spectral density of the VECSEL frequency noise when free-running (solid red line) as determined from the error signal of H\"ansch-Couillaud locking system. The free-running linewidth as determined from these measurements is shown in the legend -- see text for details. The background (dashed black line) is measured with the laser running but out of resonance with the cavity. \label{fig:figureVECSEL}}
\end{figure}

In the NIR, the VECSEL is capable of producing an output of greater than 1~W in the region of 908-915~nm and greater than 2~W between 915 and 928~nm, with a peak output of around 3~W at 925~nm (Fig.~\ref{fig:figureVECSEL}a). Moreover, the spatial mode is of a high quality, being nearly circular and having a measured M$^2$~<~1.1. This represents a clear improvement over tapered amplifier systems which, as mentioned above, suffer from poor mode quality. The practical benefits of this would include e.g. highly efficient coupling into a single-mode fibre, but here it is manifest as excellent coupling into the SHG cavity of 90\% or greater, even when using only a single spherical lens to perform the mode matching. This is advantageous as it allows nearly all of the power in the NIR to be available for frequency doubling. In comparison, although it is possible to generate a similar amount of power using an ECDL and tapered amplifier system in this regime, much lower coupling into the SHG cavity is typically achievable ($\sim$50\%), despite usually utilising more complicated optical coupling systems. Furthermore, this efficiency can vary greatly depending upon the exact tapered amplifier chip employed due to variable beam quality (M$^2$~<~1.2-1.7) and also on the operating current of the amplifier which affects the beam shape.
		
\begin{figure}[t]
\centering\includegraphics[width=\textwidth]{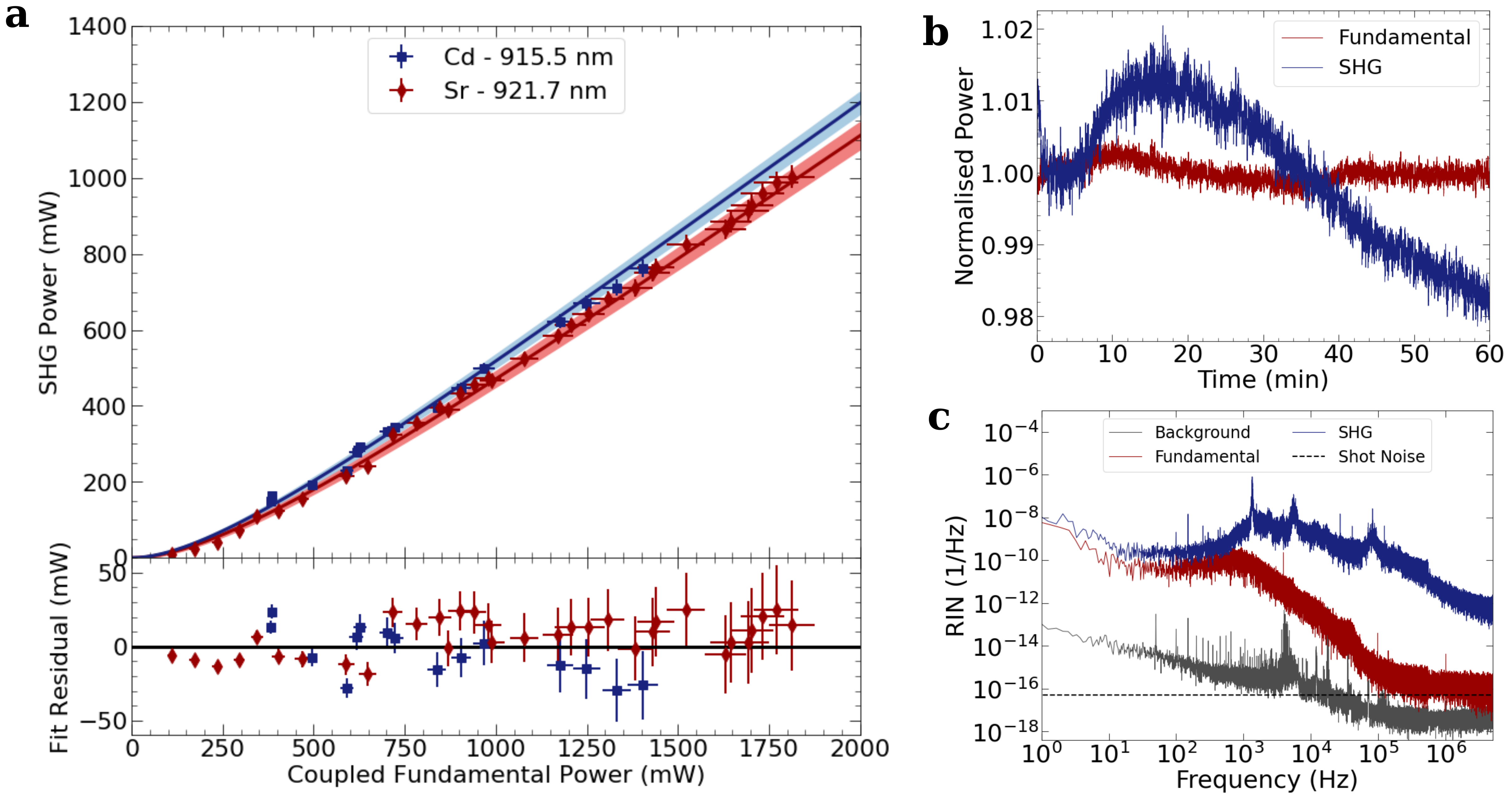}
\caption{a) SHG power as a function of the power of the fundamental light coupled into the two cavities (blue square -- Cd, red diamonds -- Sr). The data is fit by varying the linear losses of the cavity, with all other parameters fixed by theory. Errors bars represent the systematic error from the uncertainty of the calibration of the power meters ($\pm$3\%), whilst the shaded region represents the uncertainty of the fit. b) Power measurement of the SHG over the period of an hour. c) Relative intensity noise of the fundamental light (red line) incident upon the frequency-doubling cavity and of the SHG (blue line). The dashed black line is the shot-noise limit for the measurement of the fundamental (8.5~mW). The background trace (grey line) is collected with the incident light on the measurement photodiode blocked.\label{fig:figureSHG}}
\end{figure}

Fig.~\ref{fig:figureSHG}a shows the results of the doubling efficiency from the two resonant bow-tie cavities, which are capable of producing greater than 1~W at the wavelength required for the $^1S_0 \to {}^1P_1$ of strontium at 460.86~nm. Slightly less power is produced at the cadmium-specific wavelength of 457.49~nm, despite a slightly higher observed conversion efficiency, due to the reduced power of the VECSEL at this wavelength (Fig.~\ref{fig:figureVECSEL}a). Additionally, a lower coupling efficiency closer to 70\% was operational during these measurements, limited only by the use of a different set of optics. Fundamentally, there are no limitations which would prevent the possibility of also reaching 1~W at this wavelength. The data is fit using the least-squares method, with the linear losses of the cavity as the only parameter, to a theoretical prediction of the SHG based upon the cavity parameters and the measured powers~\cite{Pizzocaro_2014}. The extracted finesse values of 387$\pm$6 and 367$\pm$7, for the cadmium and strontium cavities respectively, are slightly lower than the values determined by measuring the linewidth, suggesting that even greater SHG powers may be possible with further optimization. The cavity and the laser can remain locked for long periods and the power of the SHG over one hour, as measured with an amplified, biased photodiode, shows variation contained within $\sim\pm$2\%, whilst the fluctuations of the output from the VECSEL over the same period is $\pm$0.5\% (Fig.~\ref{fig:figureSHG}b). 

The relative intensity noise (RIN) of the visible output and the NIR input are measured with a small-area, amplified photodiode with a bandwidth of $\sim$100~MHz. Fig.~\ref{fig:figureSHG}c shows that at low frequencies, the RIN of the SHG approaches the limit of the noise of the fundamental laser at around -100~dBc in the region 10-1000~Hz. At higher frequencies, the RIN of the SHG clearly exceeds that of the NIR input to the cavity, due to the limited bandwidth of the locking electronics currently employed. The performance of the VECSEL itself, however, approaches the shot-noise limit at high frequencies for the measurement power sent to the photodiode (8.5~mW).

In addition to providing a long-term, stable lock, the output of the H\"ansch-Couillaud system can be used to estimate the linewidth of the VECSEL output, when free-running and without feedback. By calibrating the slope of error signal with the measured linewidth of the cavity, it is possible to extract the power spectral density of the laser frequency fluctuations ($S_\nu\left(f\right)$), which can in turn be used to provide an estimate of the laser linewidth. There are multiple definitions of the laser linewidth and here we evaluate the linewidth based upon the phase noise spectral density ($\int_{\Delta\nu/2}^{f_{max}} S_\phi\left(f\right)df=1$, $S_\phi\left(f\right)=S_\nu\left(f\right)/f^2$), a so-called fast linewidth, and the root-mean-squared value evaluated at various timescales ($\Delta\nu_{rms}=\int_{1/t}^{f_{max}} S_\nu\left(f\right)df$). The upper limit of the integrals ($f_{max}$) is set by the maximum Fourier frequency of the collected data ($\sim$5~MHz). For the free-running laser, the measured fast linewidth is 8~kHz and $\Delta\nu_{rms}$ is 4~kHz at 100~$\mu$s and 104~kHz at 100~ms timescales. These linewidths are considerably below the natural linewidths of the transitions in question, which are $2\pi\times$30.5~MHz and $2\pi\times$90.1~MHz for the strontium and cadmium $^1S_0 \to {}^1P_1$ transitions, respectively~\cite{Nagel_2005,Xu_2004}. Moreover, this linewidth allows for easy and efficient coupling to the frequency-doubling cavity, which as well as being practically preferable, also leads to a reduction in the intensity noise of the SHG, which is a function not just of the intensity noise of the fundamental source, but also its frequency noise via the lock to the cavity. It may additionally facilitate the direct locking to higher finesse cavities for applications demanding narrower linewidths.

\section{Spectroscopy of the $^1S_0 \to {}^1P_1$ Cd Transition}
\begin{figure}[t]
\centering\includegraphics[width=\textwidth]{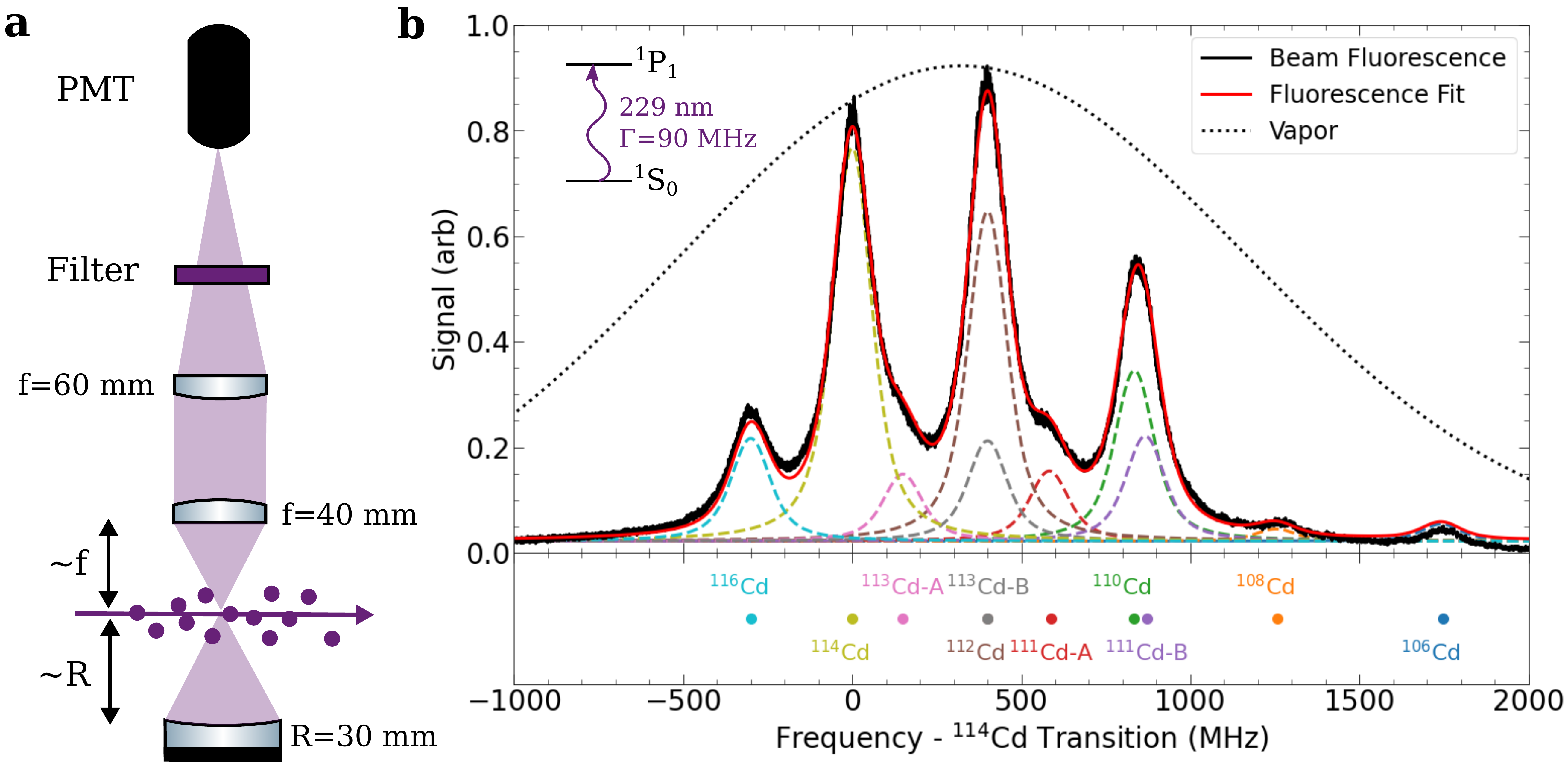}
\caption{a) The fluorescence from the cadmium beam is imaged onto a photomultiper tube. See text for more details. b) The top panel shows a sample fluorescence signal (solid black line) fit to Voigt profiles of equal width and fixed relative amplitudes given by the known natural abundances (solid red line). The dashed lines show the contributions of the individual transitions. The dotted line shows the expected Doppler-broadened profile for an ideal cadmium gas at the same temperature as the beam. The lower panel shows the mean result of multiple fits to different fluorescence data sets, with the error bars smaller than the shown data points. The separation between the points for $^{112}$Cd and $^{113}$Cd-B are also indistinguishable at this scale. For the fermionic isotopes, the labels A and B respectively refer to the $F=1/2\to F'=1/2$ and $F=1/2\to F'=3/2$ hyperfine transitions. See Table~\ref{tab:tabRef} for precise values.\label{fig:figureCd}}
\end{figure}
The suitability of this laser for atomic physics applications is demonstrated by spectroscopy performed on the 228.89~nm cadmium $^1S_0 \to {}^1P_1$ transition, using the setup shown in Fig.~\ref{fig:figureCd}a. This requires an additional stage of frequency doubling with the visible output of the first bow-tie cavity being sent to a second bow-tie cavity, this time using a beta-barium borate crystal (BBO, $\beta$-BaB$_2$O$_4$) cut at Brewster's angle as the doubling medium~\cite{Kaneda_2016,Wilson_2011,Hannig_2018}. Mode matching to this cavity is performed by a cylindrical telescope, to correct for the walk-off from the LBO cavity, and a spherical telescope and the observed coupling is approximately 70\%. Accounting for this mode-matching efficiency, we estimate that this second cavity (Agile Optic) should allow for the production of 250~mW of DUV assuming a finesse of 350 and an input-coupling mirror of 1\% transmission~\cite{Hannig_2018}.

For these spectroscopic experiments, however, the power is kept low, ensuring the delivered optical intensity is below the saturation intensity ($I/I_{sat}\sim$0.2, $I_{sat}=$0.99~W/cm$^2$), reducing the impact of power broadening and also minimising power-declining ageing effects in the crystal~\cite{Kaneda_2016}. Therefore only approximately 1~mW of collimated DUV output is sent transverse to a beam of natural cadmium in a retro-reflected configuration. By sweeping the PZT of the VECSEL with a triangular wave and keeping the two doubling cavities on resonance, it is possible to continuously tune over 4~GHz in the DUV. The fluorescence signal is measured with a photomultiplier tube and the input polarization is set to be horizontal to maximize the amount of the dipole emission collected. A lens (f=40~mm) and a curved mirror (R=30~mm), placed inside the vacuum chamber vertically offset from the atomic beam, collect the atomic fluorescence, whilst a second lens and narrow optical filter, outside the chamber, focus the light onto the PMT and remove background light (Fig.~\ref{fig:figureCd}a).

The cadmium atomic beam is based upon an oven, with a similar design to one previously used for strontium~\cite{Schioppo_2012}, with capillaries of length 10~mm and internal diameter 0.23~mm used to ensure a collimated beam. The ambient pressure when the oven is at room temperature is $\sim$10$^{-10}$~mbar. Under typical operating conditions of a temperature of around 100$^\circ$C and with a measured flow rate of $\sim$10$^{10}$~atoms/s, the divergence of the atomic beam is suitably low that it allows for different isotopes to be easily identified, as shown in Fig.~\ref{fig:figureCd}b. The observed width of the emission peaks is close to that of the natural linewidth, meaning that saturated absorption spectroscopy need not be used, with the advantage that much lower intensities may be employed. For comparison, when using an atomic vapor of cadmium at the same temperature, a single Doppler profile would cover the transitions of all the isotopes.

In addition to being coupled into the second-stage SHG cavity, a small fraction of the blue light is matched to a reference bow-tie cavity with a free spectral range of 204.2$\pm$0.8~MHz. By fitting the resonance peaks of this cavity to a parabola it provides a calibration that allows the relative frequency of the blue light with time to be determined. Furthermore, the NIR light is simultaneously coupled into a wavemeter (Bristol Instruments 621) which provides an absolute reference at an accuracy of 0.2~ppm, which corresponds to 262~MHz at the transition frequency. Finally, some of the DUV is sent to a reference photodiode, which allows for any power variations across the spectrum to be corrected for.

\begin{table}[t]
  \begin{center}
    \caption{Summary of measured isotope shift relative to the $^{114}$Cd isotope. Error values for this work sequentially represent the error from the fitting procedure and the systematic error arising from the frequency conversion -- see text for details. The two different methods from reference~\cite{Kelly_1961} are presented along with the values determined in this work. The first of these methods distinguishes neither $^{110}$Cd nor $^{112}$Cd from their respective nearby fermionic transitions, $^{111}$Cd and $^{113}$Cd, respectively, which are therefore quoted here with the same value. The second method attempts to extract the contributions of just the bosonic isotopes.}
    \label{tab:tabRef}
    \begin{tabular}{l c c r}
      \hline
      \textbf{Transition} & \textbf{Determination 1~\cite{Kelly_1961}} & \textbf{Determination 2~\cite{Kelly_1961}}  & \textbf{This work / MHz}\\
      \hline
      %\multirow
      {$^{106}$Cd} & - & - & 1747.5~$\pm$~4.3~$\pm$~9.7 \\
      $^{108}$Cd & - & - & 1256.0~$\pm$~4.3~$\pm$~7.0 \\
      $^{110}$Cd & 878~$\pm$~17 & 905~$\pm$~35 & 833.2~$\pm$~2.8~$\pm~$4.6  \\
      $^{111}$Cd - F$'$=1/2 & - & - & 587.0~$\pm$~3.1~$\pm$~3.3 \\
      $^{111}$Cd - F$'$=3/2  & 878~$\pm$~17 & - &  871.5~$\pm$~2.9~$\pm$~4.8  \\
      $^{112}$Cd  & 375~$\pm$~15  & 395~$\pm$~30 & 399.2~$\pm$~2.8~$\pm$~2.2  \\
      $^{113}$Cd - F$'$=1/2  & - & - & 150.0~$\pm$~2.9~$\pm$~0.8   \\
      $^{113}$Cd - F$'$=3/2  & 375~$\pm$~15 & - & 401.1~$\pm$~4.0~$\pm$~2.2   \\
      %$^{114}$Cd & &    \\
      $^{116}$Cd  & - & - & -299.0~$\pm$~2.6~$\pm$~1.7  \\
      
      \hline
    \end{tabular}
  \end{center}
\end{table}

In order to extract the relative frequency spacing of the different transitions, we fit the collected fluorescence signal to a sum of Voigt profiles, whose relative amplitudes are fixed based upon the known natural isotope abundances and, in the case of the fermionic isotopes $^{111}$Cd and $^{113}$Cd, the relative strength of the different hyperfine transitions. The width of the Voigt profile for the individual transitions is variable but equal for all transitions and the Lorentzian contribution set to the known natural linewidth, neglecting decoherence due to e.g. collisions. Neglecting the contributions of power broadening (6.3~MHz) and time-of-flight broadening (170~kHz), this also allows for the extraction of the atomic beam divergence from the Gaussian contribution to the Voigt profile~\cite{Schioppo_2012}, resulting in an estimated atomic beam divergence (half angle) of approximately 40~mrad. The relative frequency of each transition can be extracted from the fit results and a total of ten readings each whilst the frequency of the laser is being increased and decreased. The initial values provided to the fitting procedure of each isotope shift are selected from a uniform pseudorandom distribution with a width of 20~MHz, centred on values chosen using previous measurements~\cite{Kaneda_2016,Brickman_2007,Kelly_1961}, to reduce possible initial-value biasing effects. Using both positive and negative frequency gradients helps to average out any remaining systematic error from imperfect frequency calibration with the reference cavity.
\begin{table}[!t]
  \begin{center}
    \caption{Absolute frequency measurement for $^1S_0\to{}^1P_1$ transition of $^{114}$Cd. Error bars for this work are sequentially the statistical and systematic errors. The two values from~\cite{Burns_1956} represent values taken with two different types of lamp: a Michelson lamp; and an electrodeless lamp. Also shown are quoted values from natural cadmium.}
    \label{tab:tabAbs}
    \begin{threeparttable}
    \begin{tabular}{l l}
      \hline
      \textbf{Reference} & \textbf{Transition Frequency / MHz}\\
      \hline
      %\multirow
      $^{114}$Cd, This Work\tnote{a} & 1309864506 $\pm$ 4 $\pm$ 262  \\
      $^{114}$Cd, Michelson~\cite{Burns_1956}\tnote{b}  & 1309864420 $\pm$ 86 \\
      $^{114}$Cd, Electrodeless~\cite{Burns_1956}\tnote{b}  & 1309864600 $\pm$ 86 \\
      Natural Cd~\cite{Burns_1956}\tnote{b}  & 1309864720 $\pm$ 86 \\
      Natural Cd~\cite{NIST_Spectra}\tnote{c} & 1309864720 $\pm$ 218 \\
      Natural Cd~\cite{Moore_1958}\tnote{c}  & 1309867418 \\
      
      \hline
    \end{tabular}
    \begin{tablenotes}\footnotesize
    \item[a]Derived from NIR wavemeter.
    \item[b]Derived from wavenumber. Errors are calculated from the quoted wavelengths and their uncertainty of $\pm$0.015~pm, by calculating the refractive index at the transition wavelength ($n-1$=3.084$\times10^{-4}$)~\cite{Peck_1972}.
    \item[c]Derived from wavenumber.
    \end{tablenotes}
    \end{threeparttable}
  \end{center}
\end{table}

The determined frequency separations, relative to the $^{114}$Cd isotope transition, are presented in Fig.~\ref{fig:figureCd}b and summarized in Table~\ref{tab:tabRef}. These values are the mean, weighted by the error given by the fitting procedure, for the total of 20 data sets. The two error values respectively represent the error from fitting procedure and the systematic error which comes from the uncertainty in the frequency conversion of the temporal data. The error arising from the fitting procedure is the quadrature summation of two factors: the standard error of the mean from the fits to multiple data sets; and the mean error from all the fitting procedure of each data set, as parametrized by the covariance matrix, which is not assumed to scale inversely with repeated measurements. The error in the frequency conversion is dominated by the uncertainty in the free spectral range of the reference cavity. Errors such as the Doppler shift, detailed below, are here assumed to be common to all isotopes and therefore ignored for these relative frequency determinations. All the extracted values from the fit are consistent with the previously reported values within 3$\sigma$~\cite{Kelly_1961}.

To measure the absolute frequency of the transition of the $^{114}$Cd isotope, the wavemeter rather than the reference cavity is employed to determine the frequency axis. In this case, multiple data sets are recorded with the laser sweeping across the $^{114}$Cd transition and the traces, which are noisy due to the inherent fluctuations of the wavemeter, are fit using the same procedure as above. The mean and standard deviation of these fits give a measured value of 1309864506$\pm$4$\pm$262~MHz, where the first error is the statistical standard error of the mean and the second the systematic error. In addition to including the known accuracy of the wavemeter, the final systematic error is calculated from adding this error in quadrature to the errors arising from the first- and second-order Doppler shifts and the recoil energy. The largest of these additional systematic errors is the first-order Doppler shift, which arises from imperfect orthogonality between the atomic beam and the probe laser beam. We estimate this angle to deviate from perpendicularity by $<4$~mrad, giving an additional error of 5~MHz, meaning that the accuracy of the wavemeter remains the dominant contribution. This value of the transition frequency is consistent with references~\cite{Burns_1956,NIST_Spectra}, but not with reference~\cite{Moore_1958} (Table~\ref{tab:tabAbs}).

\section{Summary \& Outlook}
A tunable, narrow-linewidth VECSEL system has been used to generate Watt-level light in the blue region of the visible spectrum via SHG in a resonant bow-tie cavity, which is stable to the $\sim$2\% level on the timescales of one hour. The relative simplicity, narrow linewidth and good spatial mode make this system an attractive proposition for applications in both the NIR and SHG regions. The parameters obtained address the need for laser cooling and trapping of strontium and cadmium atoms, where higher powers allow for faster sample preparation times. This is an important requirement for increasing the duty cycle of experiments, which is a key target for increasing the sensitivity of many experiments.

The laser has been used as a master source in conjunction with a beam of cadmium atoms to measure the frequency of the $^{114}$Cd $^1S_0 \to {}^1P_1$ transition. New values for the isotope shifts of this transition have also been provided for all naturally occurring isotopes. To achieve a better determination of some of the individual isotopes, such as distinguishing the $^{112}$Cd transition from the $F=1/2\to F'=3/2$ $^{113}$Cd transition, it may be required to use enriched samples. When used in conjunction with the isotope shifts of other transitions of cadmium, especially when the accuracy attains the level of $\sim$kHz, such measurements could be used to make the King plots used to probe new physics~\cite{King_1963,Counts_2020}. This is therefore of particular interest for the narrow intercombination and optical clock transitions~\cite{Miyake_2019}.

The developed atomic cadmium beam itself has the potential to act as a tool for further, precision spectroscopic experiments, or to act as the atomic source for e.g. loading into a magneto-optical trap. In comparison to the vapor-based loading employed elsewhere~\cite{Yamaguchi_2019,Kaneda_2016,Brickman_2007}, this has the advantage of allowing for an improved vacuum in the trap region and improved signal-to-noise from fluorescence measurements due to reduced ambient scattering.

%\section{Assessing final manuscript length}
%OSA's Universal Manuscript Template is based on the OSA Express layout and will provide an accurate length estimate for Optics Express, Biomedical Optics Express,  Optical Materials Express, and OSA's newest title OSA Continuum. Applied Optics, JOSAA, JOSAB, Optics Letters, Optica, and Photonics Research publish articles in a two-column layout. To estimate the final page count in a two-column layout, multiply the manuscript page count (in increments of 1/4 page) by 60\%. For example, 11.5 pages in the OSA Universal Manuscript Template are roughly equivalent to 7 composed two-column pages. Note that the estimate is only an approximation, as treatment of figure sizing, equation display, and other aspects can vary greatly across manuscripts. Authors of Letters may use the legacy template for a more accurate length estimate.

%\section{Figures, tables, and supplementary materials}

\section*{Funding}
This work has been supported by the European Research Council, Grant No.772126 (TICTOCGRAV).

\section*{Acknowledgments}
We thank Jacopo Catani and Stefan Truppe for fruitful discussions. We thank Gabriele Santambrogio for the loaning of the NIR wavemeter employed in the absolute frequency measurements. We thank Fabio Corti for work on electronic systems and Tommaso Rossetti for assistance with data collection and analysis. This work has been supported by the European Research Council, Grant No.772126 (TICTOCGRAV). J.-P.P acknowledges the support of the Jenny and Antti Wihuri Foundation, the Walter Ahlstr\"om Foundation and the Finnish Foundation for Technology Promotion
%Acknowledgments should be included at the end of the document. The section title should not follow the numbering scheme of the body of the paper. Additional information crediting individuals who contributed to the work being reported, clarifying who received funding from a particular source, or other information that does not fit the criteria for the funding block may also be included; for example, ``K. Flockhart thanks the National Science Foundation for help identifying collaborators for this work.'' 

\section*{Disclosures}

%Disclosures should be listed in a separate nonnumbered section at the end of the manuscript. List the Disclosures codes identified on OSA's \href{http://www.osapublishing.org/submit/review/conflicts-interest-policy.cfm}{Conflict of Interest policy page}, as shown in the examples below:

%\medskip

%\noindent ABC: 123 Corporation (I,E,P), DEF: 456 Corporation (R,S). GHI: 789 Corporation (C).

%\medskip

%\noindent If there are no disclosures, then list ``The authors declare no conflicts of interest.''
The authors declare no conflicts of interest.

%%%%%%%%%%%%%%%%%%%%%%% References %%%%%%%%%%%%%%%%%%%%%%%%%

%Add references with BibTeX or manually.
%\cite{Zhang:14,OSA,FORSTER2007,Dean2006,testthesis,Yelin:03,Masajada:13,codeexample}

%%%%%%%%%% If using BibTeX:
\bibliography{vecselBibliography}

\end{document}